\documentclass[10pt,conference]{IEEEtran}
\IEEEoverridecommandlockouts
\usepackage{amsmath,amssymb,amsfonts}
\usepackage{algorithmic}
\usepackage{graphicx}
\usepackage{textcomp}
\usepackage{xcolor}
\usepackage{comment}
\usepackage[numbers,sort&compress]{natbib}
\def\BibTeX{{\rm B\kern-.05em{\sc i\kern-.025em b}\kern-.08em
T\kern-.1667em\lower.7ex\hbox{E}\kern-.125emX}}
\usepackage[hidelinks]{hyperref}

\usepackage[numbers]{natbib}
\usepackage[inline]{enumitem}
\usepackage{url}

\definecolor{darkgreen}{rgb}{0,0.6,0}
\newcommand{\todo}[1]{{\color{blue}$\blacksquare$~\textsf{[TODO: #1]}}}

\begin{document}

\title{Sustaining Research Software via Research Software Engineers and Professional Associations}

\author{
\IEEEauthorblockN{Jeffrey C. Carver}
\IEEEauthorblockA{\textit{Department of Computer Science} \\
\textit{University of Alabama}\\
Tuscaloosa, AL, USA \\
carver@cs.ua.edu}
\and
\IEEEauthorblockN{Ian A. Cosden}
\IEEEauthorblockA{\textit{Research Computing} \\
\textit{Princeton University}\\
Princeton, NJ, USA \\
icosden@princetone.edu}
\and
\IEEEauthorblockN{Chris Hill\todo{}}
\IEEEauthorblockA{\textit{dept. name of organization (of Aff.)} \\
\textit{name of organization (of Aff.)}\\
City, Country \\
email address or ORCID}
\and
\IEEEauthorblockN{Sandra Gesing}
\IEEEauthorblockA{\textit{Center for Research Computing} \\
\textit{University of Notre Dame)}\\
Notre Dame, IN, USA \\
http://orcid.org/0000-0002-6051-0673}
\and
\IEEEauthorblockN{Daniel S. Katz}
\IEEEauthorblockA{\textit{NCSA \& CS \& ECE \& iSchool} \\
\textit{University of Illinois}\\
Urbana, IL, USA \\
http://orcid.org/0000-0001-5934-7525}
}

\author{
\IEEEauthorblockN{Jeffrey C. Carver\IEEEauthorrefmark{1}, Ian A. Cosden\IEEEauthorrefmark{2}, Chris Hill\IEEEauthorrefmark{3}, Sandra Gesing\IEEEauthorrefmark{4}, and Daniel S. Katz\IEEEauthorrefmark{5}} 
\IEEEauthorblockA{\IEEEauthorrefmark{1}Department of Computer Science, University of Alabama, Tuscaloosa, AL, USA, Email: carver@cs.ua.edu}
\IEEEauthorblockA{\IEEEauthorrefmark{2}Research Computing, Princeton University, Princeton, NJ, USA, Email: icosden@princeton.edu}
\IEEEauthorblockA{\IEEEauthorrefmark{3}Research Computing, M.I.T, Cambridge, MA, USA, ORCID: 0000-0003-3417-9056}
\IEEEauthorblockA{\IEEEauthorrefmark{4}Center for Research Computing, University of Notre Dame, Notre Dame, IN, USA, ORCID: 0000-0002-6051-0673}
\IEEEauthorblockA{\IEEEauthorrefmark{5}NCSA \& CS \& ECE \& iSchool, University of Illinois, Urbana, IL, USA, ORCID: 0000-0001-5934-7525} 
}

\maketitle

\begin{abstract}
Research software is a class of software developed to support research. 
Today a wealth of such software is created daily in universities, government, and commercial research enterprises worldwide.
The sustainability of this software faces particular challenges due, at least in part, to the type of people who develop it.
These Research Software Engineers (RSEs) face challenges in developing and sustaining software that differ from those faced by the developers of traditional software.
As a result, professional associations have begun to provide support, advocacy, and resources for RSEs.
These benefits are critical to sustaining RSEs, especially in environments where their contributions are often undervalued and not rewarded.
This paper focuses on how professional associations, such as the United States Research Software Engineer Association (US-RSE), can provide this.
\end{abstract}


\section{Introduction and context}

We define software sustainability as the ability to ensure the usefulness of software over time, by fixing bugs, adding features, and adapting to changes in both software and hardware dependencies.
When software is sustainable, it increases reuse and decreases repetitive work.
To study sustainable software engineering practices, we have to examine both the software and the people who develop it.
While these two elements are not independent (they intersect in the study of sustainable software engineering practices), our approach is from the perspective of the people.
A ready workforce of skilled software developers is critical for sustainable software because (1) use of good software development practices increases the likelihood of sustainable software and (2) there is a need for developers to maintain software to keep it current with technology and with user needs.
Therefore, software sustainability relies on software developer sustainability.

While sustaining software developers may not be a challenge in industry, it is a challenge in domains where the economics and incentives differ, e.g., in open-source software or research projects where software is a `by-product' or intermediate product.
In this paper, we focus on \textit{research software}, where the question of sustaining software developers is particularly relevant. 
Research software is software used to generate, process, or analyze research results~\cite{uk-2014-software-study}.
This software is created daily in universities, government, and commercial research enterprises worldwide.
When researchers, a primary constituent of research software developers, develop research software, the software is frequently viewed as an intermediate step in the research process rather than as the end goal, which is the research result enabled by the software.

Recently, a new class of software developers has begun to garner significant attention for their role in the development of high-quality research software: the research software engineer (RSE). 
RSEs differ from pure researchers in that they view research software as a primary output of their work efforts. 
RSEs, by combining knowledge of a research domain and software engineering best practices, offer a long-term solution to the sustainability of research software. 

\section{Challenges}
There are two key challenges for developing and sustaining research software, both related to the developers.

\paragraph*{Challenges related to the software}
Research software is used and developed in nearly every research domain. 
Domain-expert researchers, without any formal training or education in software development, often must develop new software or add features to existing software to solve their novel research problem. 
This approach is common because it is precisely a researcher's expertise that makes them capable and qualified to solve the research problems. 

Because research software is developed in support of research, the requirements typically cannot be enumerated \textit{a priori}.
In addition, the domains supported by research software are typically  complex, requiring complex software.
Therefore, for that software to be sustainable, it must be developed well from the beginning, to allow for flexibility as needs change, and maintained well over time to keep up with current needs.
But the rate of change in computer and programming technologies makes it increasingly difficult for researchers to keep pace~\cite{Merali-sci-computing}.
The result can be disastrous: errors that go unnoticed, cause wrong results, and result in wrong scientific conclusions and retractions of highly-cited publications~\cite{science-retraction}. 

\paragraph*{Challenges related to the RSEs}
While professional RSEs can contribute to software sustainability, their status reflects the status of research software, which can be problematic. 
Universities, particularly those that focus on research, tend to value research publications and research grants.
The faculty (and sometimes research staff) who produce these publications and grants are seen as key members of the research community. 
Research software, on the other hand, is valuable to its users, who could not conduct their research otherwise, but it is typically not valued by administrators (e.g., chairs, deans), who often do not value it sufficiently to consider it to be the focus of someone's job. 
In addition, universities hire and promote very few faculty on the basis of their software work, though the number does appear to be growing.

Conversely, national labs often do value research software. 
Many staff build software as much as they produce papers.
And in industry, software can be the core of a business or a key component that enables the business to be competitive.
In both settings, developers receive more recognition and incentives.

In academia, research software developers generally lack institutional support for developing sustainable software. 
Because they lack viable career tracks and long-term funding, there is frequent turnover among these developers, who are often graduate students or postdocs. 
The turnover leads to software redevelopment rather than reuse, a lack of institutional memory about what decisions were made and why, lack of cohorts of staff to support each other, and lack of staff to cover funding shortfalls in one project with other projects.

A further challenge is the lack of formal RSE training curricula at the undergraduate level, and of professional certification recognition at postgraduate levels. 
As a result, there is limited awareness of practical challenges of developing research software and limited sharing of best practices and lessons learned. 
This situation is profoundly limiting to the emergence of a truly professional class of career RSEs.

\section{Existing solutions}
Preparing and sustaining an RSE workforce is crucial to the long-term health of research software.
Without a well-prepared pipeline and clearly defined career paths, the RSE workforce will experience turnover and setbacks resulting in lost work, duplicated effort, and unsustainable research software.
To address these challenges and provide for the sustainability of RSEs, professional associations have begun forming,
initially in the UK, and now in at least eight countries/regions, including the US (US-RSE, \href{https://us-rse.org}{us-rse.org}), our focus here.

Born from a grassroots community effort, US-RSE has taken a three-pronged approach to developing, supporting, and sustaining RSE professionals. 
\textbf{(1) Community}. US-RSE focuses on individuals such as RSE professionals, aspiring RSE professionals (e.g., students or software engineers in non-research), and RSE allies (who support the importance of the role.) 
By providing a means to connect with other like-minded individuals, US-RSE creates a peer network to support its members. 
Members can ask questions, learn from others, and share professional connections. 
In cases where an RSE is the lone software developer in a research team, this platform provides their only peer-to-peer connections. 
A community of professionals also provides an understanding of the RSE career path. 
\textbf{(2) Advocacy}. US-RSE, through web articles, videos, and formal presentations, promotes RSEs’ impact on research, highlighting the increasingly critical and valuable role RSEs serve in the long-term sustainability and performance of research software. 
\textbf{(3) Resources}. US-RSE provides a platform for sharing technical and learning opportunities to support professional development for new and existing RSEs. 
US-RSE also provides information and material designed to support the establishment and expansion of RSE groups.

\section{Strengths and Weaknesses}

The community, advocacy, and resources provided by organizations like the US-RSE have a number of strengths for the sustainability of RSEs and of research software.
Organizations like US-RSE provide the opportunity to compare university RSE structures to understand what choices are made and why~\cite{8823733}. 
Across universities we have observed higher quality software (e.g. maintainable, sustainable, and reproducible) that comes from trained professionals like RSEs. 
Organizations like the US-RSE also provide the mechanisms to connect RSEs with each other so they can more effectively advocate for their positions/funding, learn from each other, and receive training.

Further evidence of the strength of RSE associations like US-RSE is rapid membership growth, a clear indication of the importance and need of such communities. 
The recent increase in the use of term ``RSE,'' both formally (in job descriptions) and informally (identifying with the role), has been a clear success of the RSE movement. 
Including professional RSEs in research projects improves the long-term maintainability, sustainability, and performance of software products, a clear benefit for everyone involved including researchers, students,  administrators, and funding agencies.

The primary weakness of associations like the US-RSE is the lack of formal influence to improve the landscape for RSEs.
Many positive impacts described above resulted from informal connections created through these associations.
Open questions still exist concerning academic contracts/funding, credit for software contributions, and building a formal RSE career path. 
Significant work remains to be done in understanding the best way to prepare an RSE pipeline and training/education. As more research software success stories stem from RSE involvement, coherent community organizations such as US-RSE will play a vital role in sharing and increasing the awareness within the larger research ecosystem. 
  
\bibliographystyle{IEEEtran}
\bibliography{bokss-2021.bib}

\end{document}